\documentclass[a4paper,fleqn,usenatbib]{mnras}
\pdfoutput=1
\usepackage{txfonts}
\usepackage[T1]{fontenc}
\usepackage{ae,aecompl}

\usepackage{graphicx}	% Including figure files
\usepackage{float}
\usepackage{epsfig}
\usepackage{epstopdf}
\usepackage{color}

\usepackage{url}

\newcommand{\gal}{NGC\,448}
\newcommand{\kms}{$\textrm{km~s$^{-1}$}$}
 
\newcommand{\Hb}{H$\beta$}

\newcommand{\nb}{\textsc{nbursts}}
\newcommand{\atlas}{ATLAS\textsuperscript{3D}}

\title[Counter-rotation in \gal]{Stellar counter-rotation in lenticular
galaxy NGC~448}

\author[I. Yu. Katkov et al.]{
Ivan Yu. Katkov$^{1}$\thanks{E-mail: katkov@sai.msu.ru (IYK)}
Olga K. Sil'chenko,$^{1,2}$ 
Igor V. Chilingarian,$^{3,1}$
Roman I. Uklein$^{4}$
\newauthor
and Oleg V. Egorov$^{1}$
\\
$^{1}$Sternberg Astronomical Institute, M.V. Lomonosov Moscow State University, Universitetskiy pr., 13,  Moscow, 119992, Russia\\
$^{2}$Isaac Newton Institute of Chile, Moscow Branch\\
$^{3}$Smithsonian Astrophysical Observatory, Harvard-Smithsonian Center for Astrophysics, 60 Garden St. MS09, Cambridge, MA 02138 USA\\
$^{4}$Special Astrophysical Observatory, Nizhniy Arkhyz, Karachai-Cherkessia 369167, Russia\\
}

\date{Accepted XXX. Received YYY; in original form ZZZ}

\pubyear{2016}

\begin{document}
\label{firstpage}
\pagerange{\pageref{firstpage}--\pageref{lastpage}}
\maketitle

% Abstract of the paper
\begin{abstract}

The counter-rotation phenomenon in disc galaxies directly indicates a complex
galaxy assembly history which is crucial for our understanding of galaxy
physics. Here we present the complex data analysis for a
lenticular galaxy NGC~448, which has been recently suspected to host a
counter-rotating stellar component. We collected deep long-slit spectroscopic
observations using the Russian 6-m telescope and performed
the photometric decomposition of Sloan Digital Sky Survey (SDSS) archival images.
We exploited (i) a non-parametric approach in order to recover stellar line-of-sight
velocity distributions and (ii) a parametric spectral decomposition technique 
in order to disentangle stellar population properties of both main and counter-rotating
stellar discs. Our spectral decomposition stays in perfect agreement with the 
photometric analysis. The counter-rotating component contributes $\approx$30 per cent
to the total galaxy light. We estimated its stellar mass to be
$9.0^{+2.7}_{-1.8}\cdot10^{9}M_\odot$. The radial scale length of
counter-rotating disc is $\approx$3 times smaller than that of the main disc.
Both discs harbour old stars but the counter-rotating components reveals
a detectable negative age gradient that might suggest an extended inside-out formation 
during $3\dots4$ Gyrs. The counter-rotating disc hosts more metal-rich stars and
possesses a shallower metallicity gradient with respect to the main disc. Our findings
rule out cosmological filaments as a source of external accretion which is
considered as a potential mechanism of the counter-rotating component formation in
NGC~448, and favour the satellite merger event with the consequent slow gas
accretion.

\end{abstract}

% Select between one and six entries from the list of approved keywords.
% Don't make up new ones.

%%%%%%%%%%%%%%%%%%%%%%%%%%%%%%%%%%%%%%%%%%%%%%%%%%

%%%%%%%%%%%%%%%%% BODY OF PAPER %%%%%%%%%%%%%%%%%%

\section{Introduction}

The kinematical appearance, as well as the stellar population properties keep
a fossil record about the process of galaxy assembly which is crucial for our
understanding of galaxy physics. Most disc galaxies possess regular rotation
but systems with peculiar kinematics are also observed. \citet{Rubin1994}
proposed a term ``multi-spin galaxies'' for objects possessing embedded
kinematically decoupled components, the angular momentum of which differs from
that of the host galaxy. The class of multi-spin galaxies includes objects with
various kinematically distinct components containing gas and/or stars having
different spatial extent and inclination with respect to the main galaxy disc:
inner polar discs, extended polar rings/discs, moderately inclined rings,
kinematically decoupled cores, and extended counter-rotating components.

Studies of multi-spin components shed the light on the process of gas accretion 
and merging history which are thought to be responsible for shaping disc galaxies.
The \textit{stellar} counter-rotation confined to the main galactic plane can be
considered as a final product of evolution of externally acquired gas that has
been processed through at least two major steps: (i) settling of the gas into the
equatorial galactic plane due to, for instance, dynamical influence of the
gravitational potential of the main disc \citep{Tohline1982}; (ii) and subsequent star
formation \citep{Pizzella2004}. Potentially, studies of counter-rotating
components may allow us to date accretion events and the following star
formation that leads to understanding of the galaxy assembly.

The counter-rotation phenomena were deeply studied by means of numerical
simulations, \citet{Thakar1996, Thakar1998} investigated various mechanisms
which are able to produce a counter-rotating gas component, such as episodic and
continuous gas infall or merger with a gas-rich dwarf satellite. In their
simulations, only small counter-rotating stellar discs with radial density
profiles different from exponential were generated.  Binary major
mergers of giant galaxies usually destroy discs and form ellipticals
\citep{Barnes_Hernquist1991}. However, a strictly coplanar merger of two gas-rich
giant progenitor galaxies is able to build up a massive counter-rotating disc
\citep{Puerari_Pfenniger_2001, Crocker2009}.  Otherwise, recent numerical
cosmological simulation by \citet{Algorry2014} naturally predicts the formation
of counter-rotating discs as a consequence of gas accretion from two distinct
filamentary structures.

Presently, several dozens of galaxies with counter-rotating components are known
\citep{Galletta1996, Corsini_review}. The majority of them have been only
suspected to host counter-rotating components. Particular progress in the
studies of counter-rotating discs has been achieved upon
development of the spectral decomposition approaches. For the first time,
such an approach was applied by \citet{Chil_Mieske2011} in order to recover a 
spectrum of an ultra-compact dwarf galaxy contaminated by the host galaxy light.
Independently, \citet{coccato_n5719} firstly proposed the spectral
decomposition approach based on the spectral pixel fitting technique \textsc
{ppxf} \citep{ppxf} for the investigation of the counter-rotation in
NGC~5719. The spectral decomposition allows one to disentangle
contributions of both counter-rotating and main components to the observed
spectrum and to determine stellar population properties and, therefore, constrain
their formation mechanisms. \citet{coccato_n5719} simultaneously
measured kinematics and stellar population properties and found that less massive
counter-rotating component is younger, less metal abundant and $\alpha$-element
enhanced with respect to the main stellar disc. Later, Coccato~et~el. and other teams,
including ours, analysed spectra of a few counter-rotating galaxies by
using the same or similar approaches. The disc galaxies with counter-rotation,
NGC~3593, NGC~4138, NGC~4191, NGC~4550, NGC~5719, and IC~719, have been studied
in this manner \citep{coccato_n3593_n4550, Pizzella_n4138, coccato_n5719,
Johnston_n4550, coccato_n4191,ic719}. The results of these studies indicate
that the counter-rotating components in all those objects being less
massive than the main stellar discs, have younger stellar populations than the
main discs while their metallicities and $\alpha$-element abundances
can deviate in both directions. Those findings are in agreement with the formation
scenario where counter-rotating stars emerge from the externally accreted gas
\citep{Pizzella2004}.

In this paper we expand the sample of well-studied counter-rotating galaxies by
one more object. We present an analysis of new deep long-slit spectroscopic data
for a counter-rotating lenticular galaxy NGC~448.

% FGC227 counter-rotatingns thick-disc Yoachim \& Dalcanton (2008)

\section{\gal: general description}

\begin{figure}
\centering
\includegraphics[width=0.9\columnwidth]{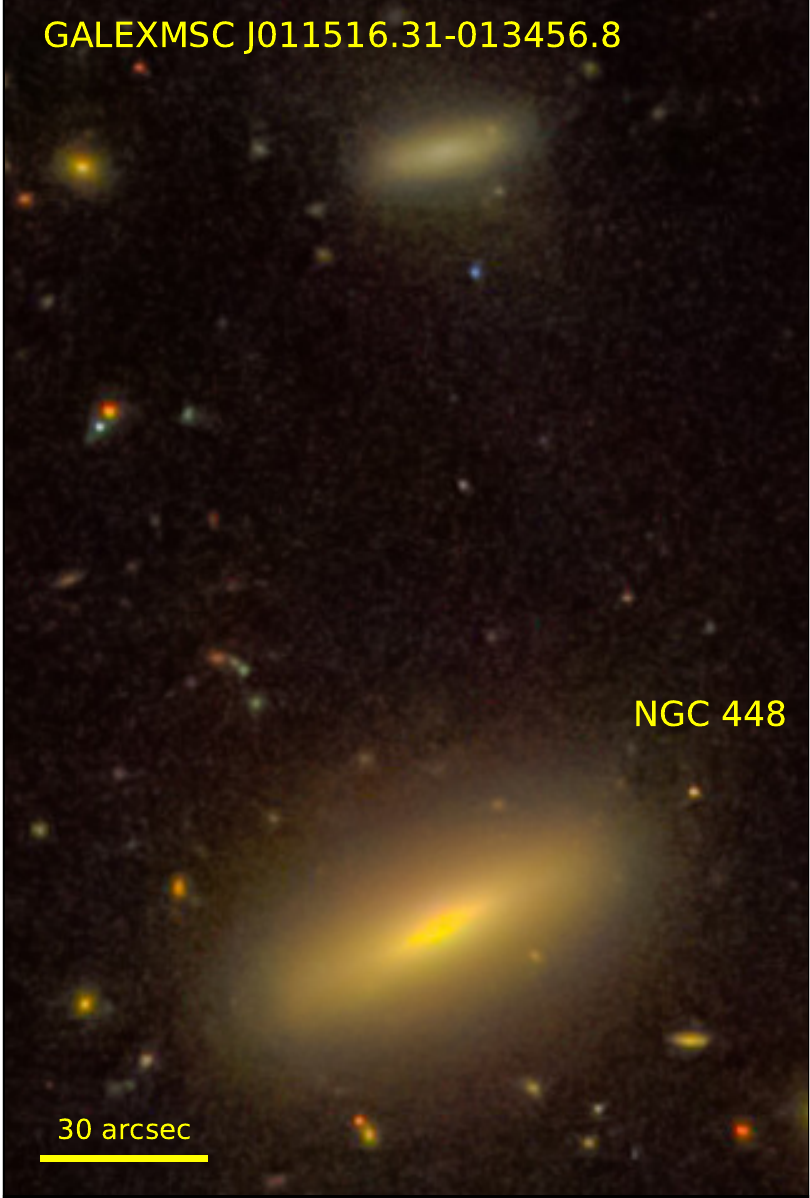}
\caption{Coloured SDSS (DR12 \citet{sdss_dr12}) image of NGC~448 and its
satellite GALEXMSC J011516.31-013456.8. North is up, east is left. The image was
extracted using the cutout image service\protect\footnotemark.}
\label{fig_sdss_image}
\end{figure}

NGC~448 is a lenticular galaxy located at a distance of 29.5 Mpc
\citep{atlas3d_i} with the total luminosity $M_B=-19.17$~mag (according to the
HyperLeda database\footnote{\url{http://leda.univ-lyon1.fr/}} \citep
{hyperleda}) and $M_K=-23.02$~mag \citep{atlas3d_i}. In the RC3 \citep{rc3}
NGC~448 is classified as S0\textasciicircum- edge-on. Nevertheless, the direct
SDSS image (Fig.~\ref{fig_sdss_image}) does not explicitly indicate a strongly
edge-on galaxy orientation.

% \begin{figure}
% \labellist
% \pinlabel {\textcolor{yellow}{NGC 448}} at 80 150 
% \pinlabel {\textcolor{yellow}{GALEXMSC J011516.31-013456.8}} at 150
% 380
% \endlabellist
% \centering
% \includegraphics[width=0.4\textwidth]{pics/sdss_n448}
% \caption{Coloured SDSS image of NGC~448 and its satellite GALEXMSC J011516.31-013456.8.
% The orientation is the north to the up, east to the left.}
% \label{fig_sdss_image}
% \end{figure}

\footnotetext{SDSS image cutout tool 
\url{http://skyserver.sdss.org/dr12/en/help/docs/api.aspx\#cutout}}

% \begin{figure}
% \centering
% \includegraphics[width=0.45\textwidth]{pics/sdss_n448_only.jpg}
% \caption{Coloured SDSS image of NGC~448 extracted through cutout image
% service\protect\footnotemark.}\label{fig_sdss_image}

% \end{figure}
% \footnotetext{SDSS image cutout tool \url{http://skyserver.sdss.org/dr12/en/help/docs/api.aspx\#cutout}}

NGC~448 was included into the \atlas\ integral-field spectroscopic survey 
\citep{atlas3d_i}. For the first time, NGC~448 was noticed as a galaxy
possessing two large-scale counter-rotating disc-like stellar components by
\citet{atlas3d_ii} basing on \atlas\ stellar line-of-sight velocity and velocity
dispersion maps which indicate a counter-rotating core and double peak features,
correspondingly.
 
NGC~448 has a companion galaxy 2.6~arcmin to the North which corresponds
to the projected separation of 22~kpc for the adopted galaxy distance of 29.5~Mpc.
According to the NASA/IPAC Extragalactic Database
(NED)\footnote{\url{http://ned.ipac.caltech.edu/}}, this galaxy is 
cross-identified with GALEXMSC J011516.31-013456.8, which radial
velocity $v_r=22105$ \kms\ was determined in the CAIRNS (Cluster and Infall
Region Nearby Survey) project \citep{cairns}. However, extremely deep optical
imaging with the MegaCam camera at the Canada-France-Hawaii Telescope (CFHT)
unambiguously shows a tidal interaction between NGC~448 and a disturbed
companion (see Fig.~16 in \citet{matlas}\footnote{The deep image
obtained with the MegaCam camera indicating the tidal interaction between NGC~448
and companion can be seen at the URL: 
\url{http://irfu.cea.fr/Projets/matlas/public/Atlas3D/NGC0448_meg.html}}).
One can see a faint tidal tail around GALEXMSC J011516.31-013456.8 even at
the contrast-enhanced SDSS image (Fig.~\ref{fig_sdss_image}). This points to
the wrong redshift measurement for GALEXMSC J011516.31-013456.8 and suggests
the physical interaction between this galaxy and NGC~448.

Investigation of stellar population properties by using optical 
long-slit spectra revealed the old stellar population ($\sim$9 Gyr) and slightly 
sub-solar metallicity in NGC~448 \citep{caldwell2003}. Similar values were
determined by \citet{atlas3d_xxx} based on the SAURON Lick index
measurements within one effective radius ($R_{eff}=11.2''$): 
$T_{SSP}=8.0\pm1.5$ Gyr, $Z_{SSP}=-0.21\pm0.05$ dex, [$\alpha$/Fe]$=0.1\pm0.06$~dex. 
Observations of the CO emission provides only upper limits on the mass
of molecular hydrogen, $M($H$_2)<5.5\cdot10^7$ M$_\odot$, \citep{atlas3d_iv}. We
do not find any data on the H{\sc i} content of NGC~448.

\section{Data analysis}\label{Obs}
%In this section we describe the observations and the data analysis.
\subsection{Observations and data reduction}

We obtained new deep spectroscopic data for NGC~448 on October 25, 2013 by using
the SCORPIO universal spectrograph \citep{scorpio} mounted at the prime focus of
the Russian 6-m BTA telescope operated by the Special Astrophysical Observatory.
We used the long-slit mode with the slit width of 1.0~arcsec placed along the major
axis of the galaxy. The volume phase holographic grism VPHG2300G gives the
spectral resolving power $R\approx2000$ (instrumental velocity dispersion of
$\sigma_{inst}\approx 65$ \kms) in the wavelength range 4800$\dots$5600 \AA\ which
includes strong absorption (Mg$b$, Fe$5270$, Fe$5335$, etc.) and emission lines (\Hb, 
[O{\sc iii}], [N{\sc i}]). We collected 16 exposures of 15~min each (4
hours in total) under good atmospheric transparency with the average seeing of
1.5~arcsec FWHM. The detector CCD EEV42-40 (2048$\times$2048 pixels)
provides the spectral sampling of 0.37~\AA\ pix$^{-1}$ and the slit plate scale
of 0.357~arcsec~pix$^{-1}$ in the $1\times2$ binning mode. In addition to the
science spectra, we obtained internal flat field, arcs (He-Ne-Ar) and twilight 
spectra.

We reduced our spectroscopic data with our own \textsc{idl}-based pipeline which
consists of standard procedures such as bias subtraction, flat fielding, cosmic
ray hit rejection by using Laplacian filtering technique L.A.Cosmic 
\citep{lacosmic}, the wavelength calibration, the sky background subtraction and
correction for the spectral sensitivity inhomogeneity. In order to build the wavelength
solution, we fitted the arc line positions using bivariate polynomial of the 5th
degree along dispersion and the 4th degree across dispersion in order to take into
account the slit image curvature. A typical value of the RMS in every line is 0.05~\AA. Due to
optical distortions, the spectral line spread function (LSF) of the spectrograph
varies along the slit as well as within the wavelength range. The LSF variations
affect the night-sky spectrum which is derived by using outer areas of the frame
where LSF shape differs from that in the regions of the galaxy close to the
slit center. In order to take into account LSF variations, we used the reconstruction
method in the Fourier space described in \citet{katkov2014_ilgpop_skysubtr};
a detailed discussion on the sky subtraction for a changing LSF is provided in
\citet{chil_6340_skysubtr,skysubtr_adass_proc2011}. Finally, we linearized
the spectra and integrated all separate exposures controlling the position of
the bright galaxy center in order to take into account the atmospheric refraction. We computed error frames
from the photon statistics and proceeded them through all reduction step. We exploited
the twilight spectrum in order to extract the LSF along the slit by fitting it
against a high-resolution solar spectrum.

\begin{figure}
\centering
\includegraphics[width=0.5\textwidth]{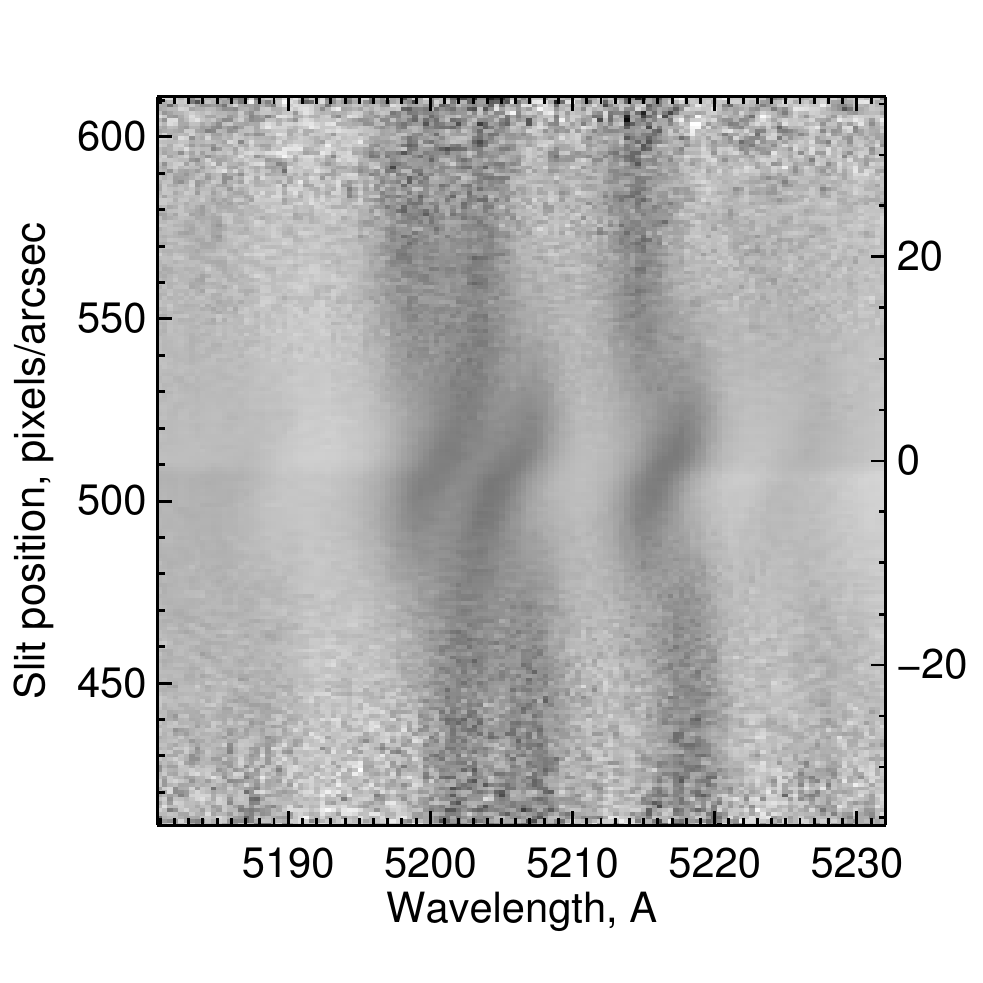}

\caption{Fragment of the long-slit spectrum around the Mg$b$ absorption feature. 
In order to achieve a better contrast, the spectrum was normalized by the average light
profile. Two kinematically distinguished components are clearly seen.}
\label{fig_longslit_fragment}

\end{figure}

Fig.~\ref{fig_longslit_fragment} demonstrates a fragment of the co-added reduced
spectrum of NGC~448 near the strong Mg$b$ absorption feature. The spectrum is
normalized by mean values in every row along the slit for presentation purposes
in order to achieve the higher contrast. The complex velocity structure of the
absorption lines which corresponds to kinematically separated stellar components
is clearly seen.

\subsection{The Non-parametric LOSVD Reconstruction}

A visual inspection of strong absorption lines in the spectrum (see
Fig.~\ref{fig_longslit_fragment}) reveals a complex, multi-component structure
of the line-of-sight velocity distribution (LOSVD) of stars which cannot be
successfully described by a single Gaussian or a Gaussian-Hermite function 
\citep{gausshermite}.

We applied a non-parametric reconstruction approach in order to determine the stellar LOSVD of
\gal. First, we adaptively bin the long-slit spectrum in the spatial
direction in order to reach a minimal signal-to-noise ratio of $S/N=30$ per bin
per spectral pixel in the middle of the spectral range. After that, we
determine a template spectrum  by using the \nb\ package
\citep{nbursts_a,nbursts_b}. This package implements a pixel-to-pixel $\chi^2$
minimization fitting algorithm where an observed spectrum is approximated by
a stellar population model broadened with a parametric LOSVD (the Gaussi-Hermite shape
is used at this step) and multiplied by some polynomial continuum in order to take into
account dust attenuation and/or possible flux calibration errors in both
observations and models. We used a grid of {\sc pegase.hr}  high resolution
simple stellar population (SSP) models \citep{pegasehr} based on
the ELODIE3.1 empirical stellar library \citep{elodie3.1}, the Salpeter initial mass
function and pre-convolved with the SCORPIO LSF determined from a twilight
spectrum  as explained above. Then, we used the
stellar population model broadened with the LSF only (without the LOSVD) as a template spectrum for
the non-parametric LOSVD reconstruction. The reconstuction technique does not require any
\textit{a priori} knowledge about the LOSVD shape and searches the solution of the convolution
problem as a linear inverse ill-conditioned problem by using a smoothing
regularization. For more details, see \citet{ic719} and \citet{Katkov_2011n524}
where we applied the same approach to recover a counter-rotating stellar disc in
the lenticular galaxy IC~719 and a complex LOSVD in NGC~524.

\begin{figure}
\centering
\includegraphics[width=0.45\textwidth]{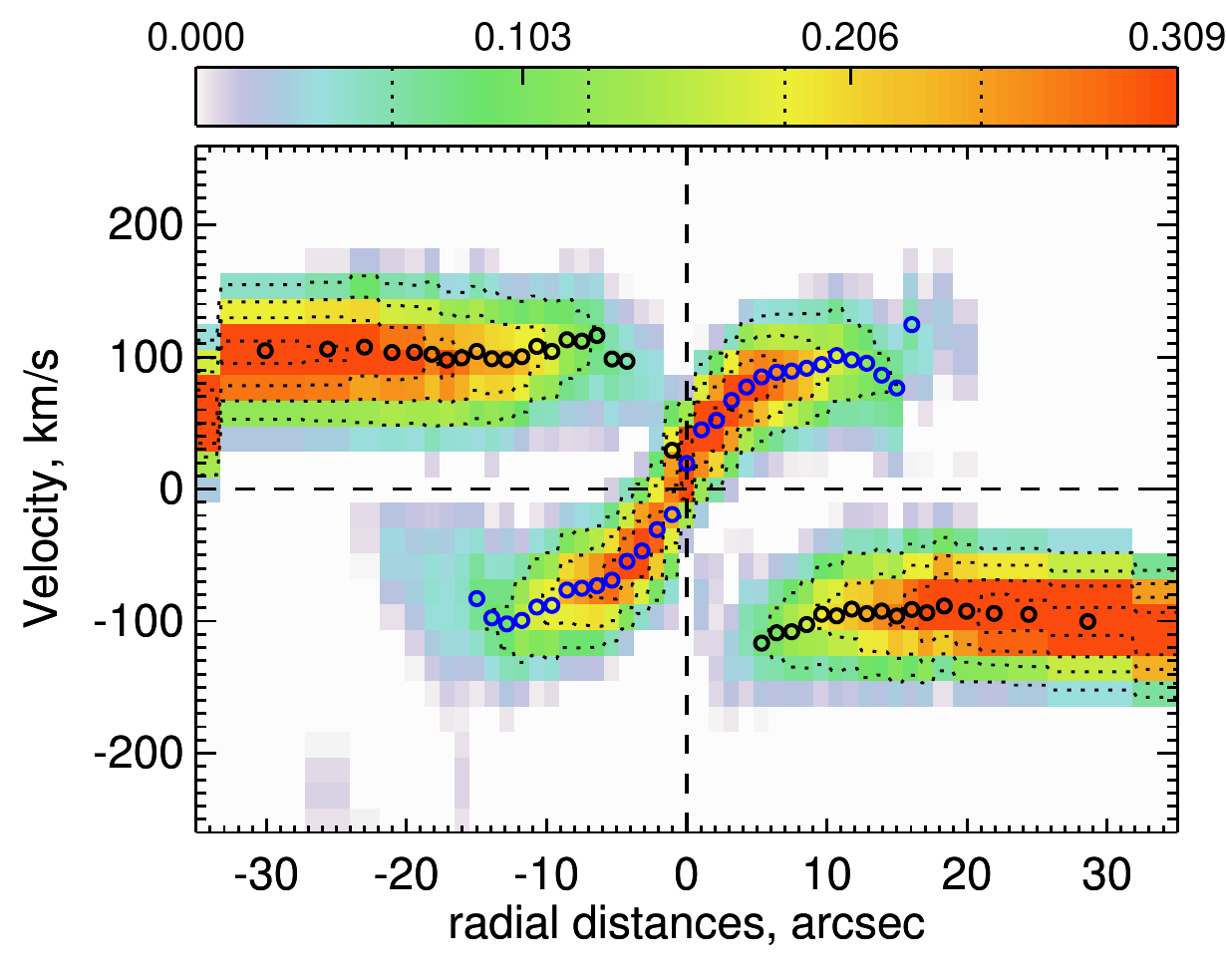}

\caption{Our result of the non-parametric reconstruction of the stellar LOSVD presented as a
position-velocity diagram. Black and blue circles show Gaussian peak positions for
a double Gaussian decomposition.}\label{fig_nonpar_losvd}

\end{figure}

Fig.~\ref{fig_nonpar_losvd} shows the reconstructed non-parametric stellar LOSVD of NGC~448
for each spatial bin. Again a complex two-component peaked structure is clearly
visible. Then we fitted a non-parametric LOSVD by a combination of two Gaussians in every bin
and estimated the line-of-sight velocities and stellar velocity dispersions for
the two kinematically distinct components.

\subsection{The Parametric Spectral Decomposition}

In order to study stellar population properties of the counter-rotating and
main stellar components, we separated their contributions to the integrated
spectrum by using the two-component mode of the \nb\ package. In general, we fitted
an observed spectrum by a linear combination of two SSP templates, each
characterized by its own age and metallicity, and broadened by different
Gaussian-shaped LOSVDs. The multiplicative polynomial continuum was the same for
both components. The Gaussian parameters of the previous non-parametric LOSVD
decomposition were used as an initial guess for the kinematics of components.
Finally, the free parameters of our model are LOS velocity, velocity dispersion,
age, metallicity for every component, relative weights and the 15th order the
polynomial continuum, hence, 24 parameters in total. Thanks to the high signal-to-noise
ratio and relatively high spectral resolution, the fitting procedure was stable
enough so we did not have to fix any parameters of the model. In order to
estimate the parameter uncertainties, we ran a Monte-Carlo simulations
for a hundred realization of synthetic spectra for each spatial bin
which were created by adding  a random noise corresponding to the signal-noise
ratio in the bin to the best-fitting model.

\begin{figure}
\centering
\includegraphics[width=0.45\textwidth]{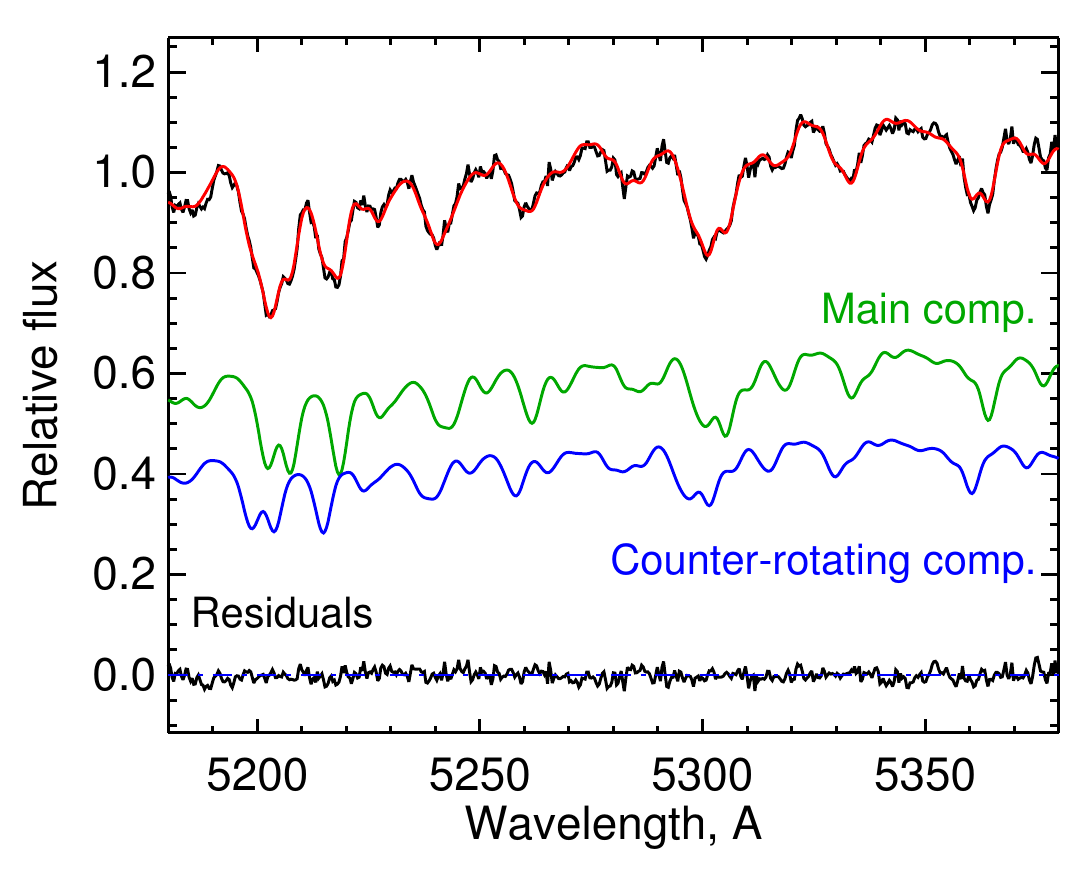}

\caption{Fragment of the spectrum at a radius $r=-12$~arcsec. The red line corresponds
to the best-fitting model overplotted on top of the observed spectrum. Green and blue lines
show the main disc and the counter-rotating component, correspondingly. }
\label{fig_sp_decomp_spectrum}
\end{figure}

\begin{figure*}
\centering
\includegraphics[clip,trim={0.1cm 0 0.1cm 0},width=0.32\textwidth]{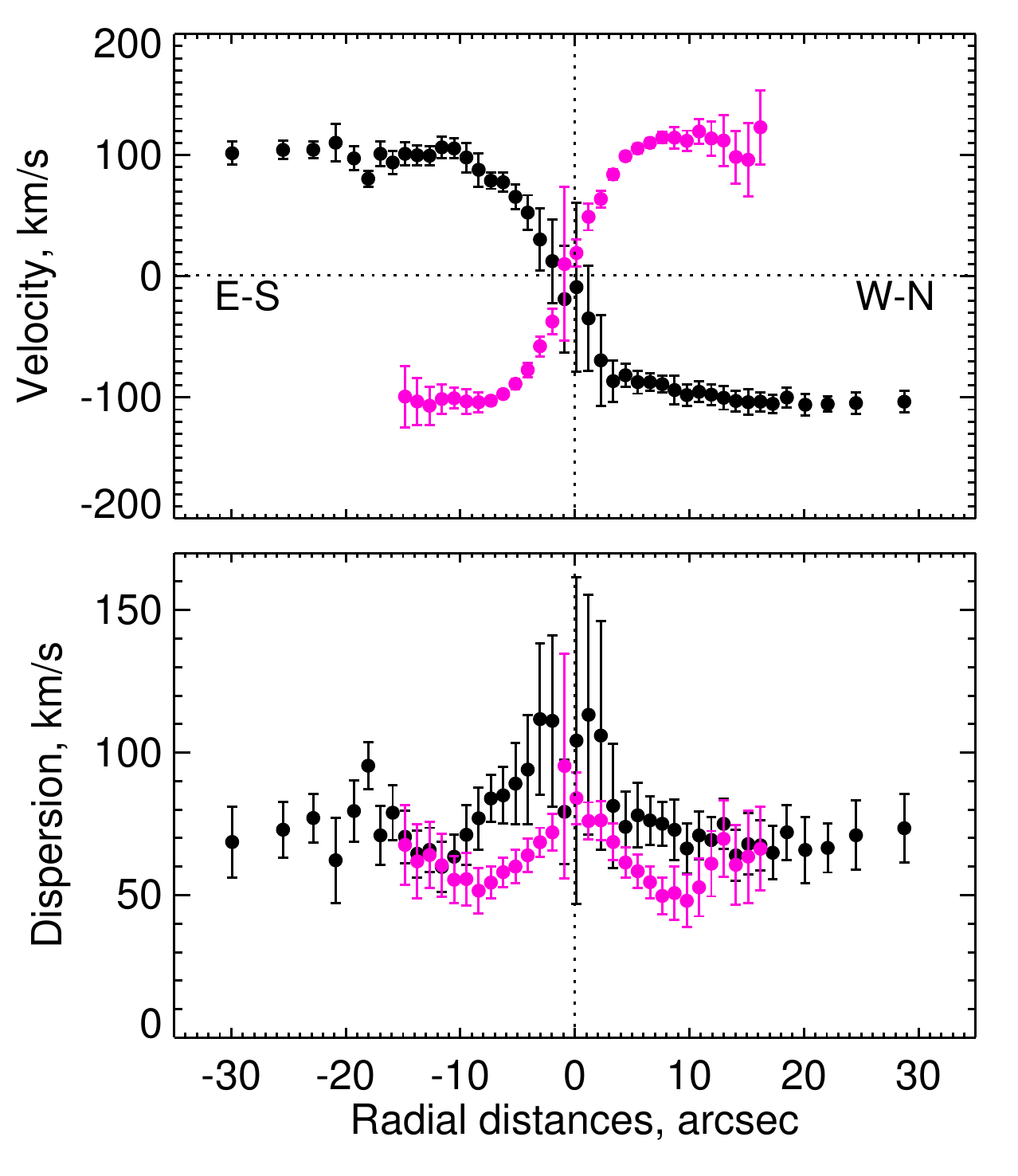} 
\includegraphics[clip,trim={0.1cm 0 0.1cm 0},width=0.32\textwidth]{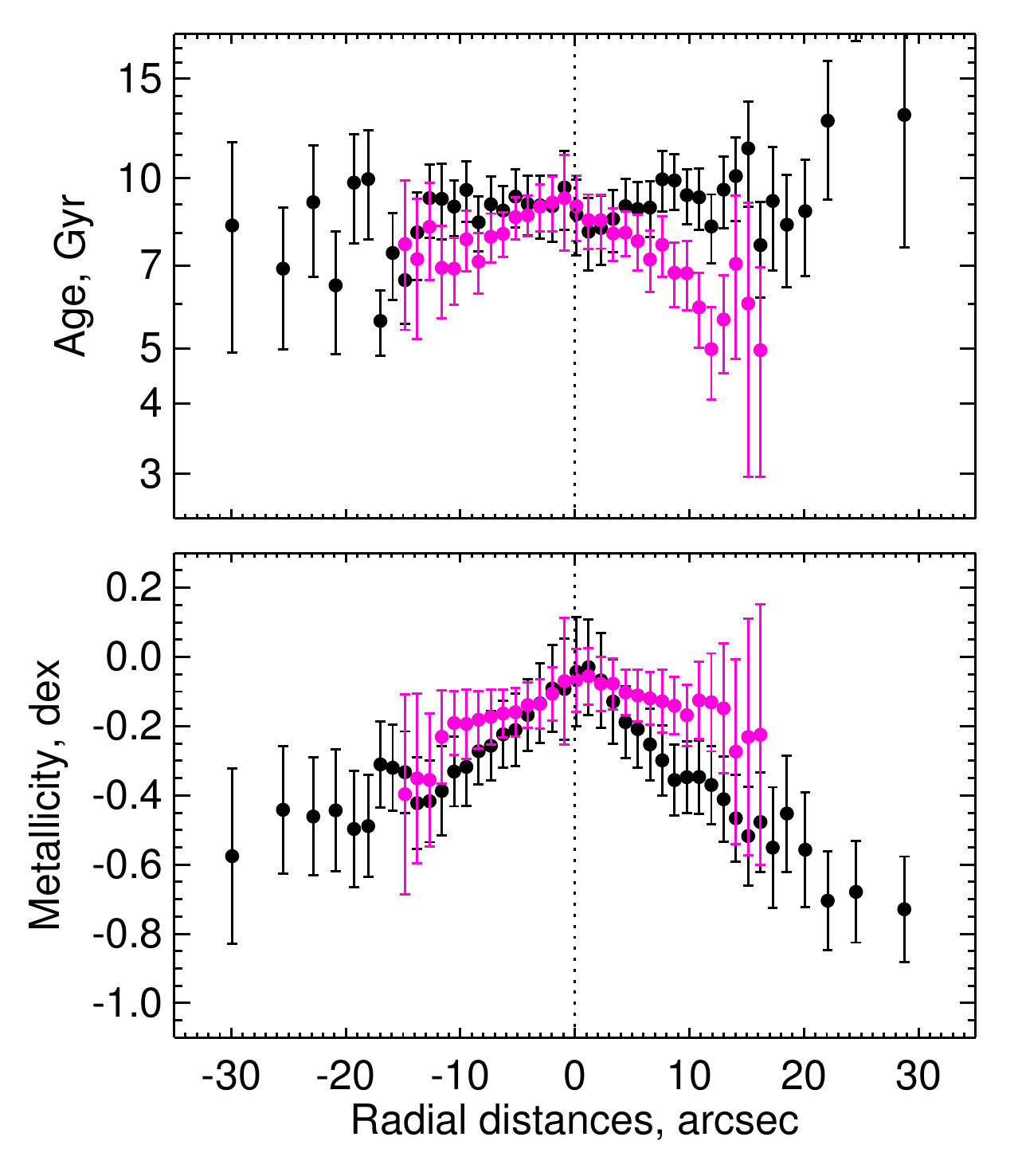}
\includegraphics[clip,trim={0.1cm 0 0.1cm 0},width=0.32\textwidth]{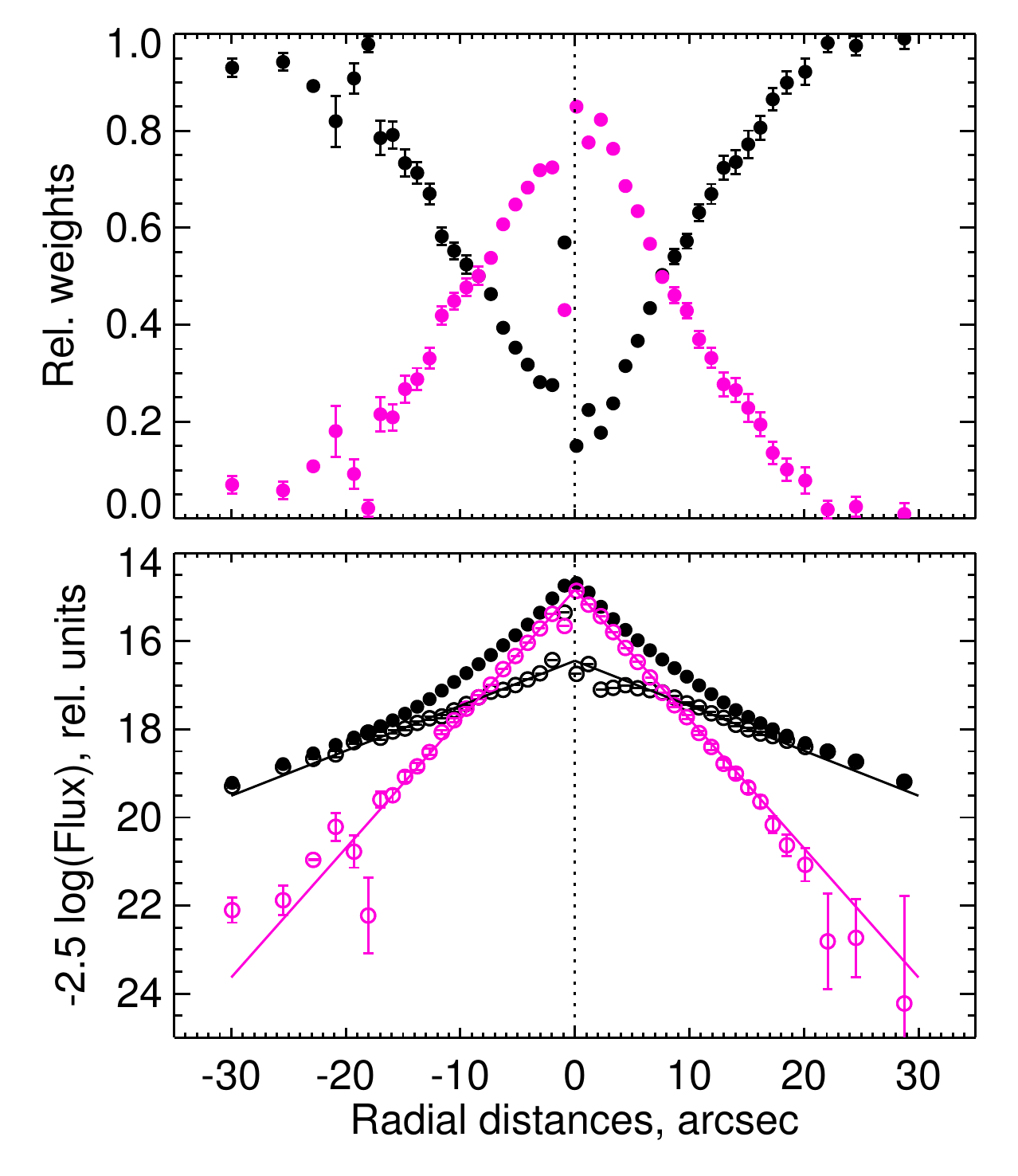}
\caption{The spectral decomposition results for \gal. Black symbols
correspond to the main stellar component, magenta symbols to the
counter-rotating disc. Left panels show radial profiles of the line-of-sights
velocity (top) and the stellar velocity dispersion (bottom). Middle panels:
SSP equivalent stellar population ages (top) and metallicities (bottom).
Right panels: relative contribution of every component to the integrated
spectrum (top); the bottom panel shows the light profile which was obtained by
integrating the spectra along the slit (filled symbols) and
light profiles of each component (open symbols).}\label{fig_sp_decomp_profiles}
\end{figure*}

Fig.~\ref{fig_sp_decomp_spectrum} shows an example of a spectrum with the
overplotted best-fitting model in a region  where the complex two-component
structure of absorption lines is clearly seen. The decomposed radial profiles
of the kinematics and stellar populations are shown in the 
Fig.~\ref{fig_sp_decomp_profiles}.

\subsection{The Photometric Decomposition}

As a complementary approach to the spectroscopic decomposition, we performed a
two-dimensional (2D) decomposition of a SDSS $g$-band image of NGC~448. We used
the $g$-band in order to compare the photometric decomposition results with
those obtain from the spectra since the spectral range of our observations is
close to the SDSS $g$-band.

In order to construct a photometric model of the galaxy, we used the
\textsc{galfit} \citep{galfit} software. An accurate treatment of uncertainties,
both for the galaxy image and for the point spread function (PSF), is required
in order to obtain reliable component parameters of the photometric model when using
\textsc{galfit}. We took into account the image uncertainties following the SDSS
documentation\footnote{\url{ http://data.sdss3.org/datamodel/files/BOSS_PHOTOOBJ
/frames/RERUN/RUN/CAMCOL/frame.html}\label{sdss_documentation}}. The PSF 
function was determined by the \textsc{SExtractor/PSFEx} packages
\citep{sextractor, psfex}. We masked background/foreground interlopers (stars,
galaxies) superimposed on the galaxy by using the fitting residuals map and then
repeated the fitting process.

The field around NGC~448 is crowded because of Galactic foreground stars. 
That could potentially lead to the imperfect sky subtraction and, consequently, bias
the photometric decomposition. We re-analyzed the sky unsubtracted SDSS image
(constructed according to the published documentation\footnotemark[\value{footnote}]) and estimated
the sky level by using azimuthally averaged surface brightness profile determined by
fitting ellipses with constant ellipticity and position angle at large radii
from the galaxy center (150--200 arcsec). We found that our sky background estimate
differs from the SDSS pipeline value\footnote{The SDSS  pipeline produces
bivariate sky map. To compare it with our sky level estimate we averaged this
map within the galaxy area, which is used for the photometric decomposition.}  by 0.2~per~cent which stays within uncertainties of our background estimate.

\begin{table}
\caption{Statistical tests of photometric models.}
\label{tbl_model_tests}
\centering
\begin{tabular}{p{4.55cm}p{0.7cm}p{0.7cm}p{0.7cm}}
\hline
Model & $\chi^2/DOF$ & $\Delta$BIC & $\Delta$AIC \\
\hline
2 S\'ersic + 2 Exp. disc  & 1.058 & 0.0  & 0.0 \\
2 S\'ersic + Exp. disc + Edge-on disc     & 1.068 & 656.1  & 665.2 \\
2 S\'ersic + Exp. disc & 1.146 & 5984.7 & 6048.8 \\
\hline	
\end{tabular}
\end{table}

We performed numerous tests and found that the galaxy image can be successfully
decomposed into four components: a S\'ersic nuclear core, two exponential discs
and a halo described also by a S\'ersic function. 

Beside this model, we tested other photometric decompositions: (i) 2
S\'ersic + exponential disc + edge-on disc; and (ii) 2 S\'ersic + exponential
disc. In order to choose between models and to measure quantitatively how well
those models fit the same data, we compared normalized $\chi^2$ values, the
Akaike and Bayesian information criteria (AIC, \citet{AIC}; BIC, \citet{BIC})
which we present in the Table~\ref{tbl_model_tests}. The model with the lowest
AIC or BIC is preferred, though a difference $\Delta$AIC or $\Delta$BIC of at
least $\sim$6 is usually required before one model can be deemed clearly superior
\citep{IMFIT}.

\begin{figure*}
\centering
\includegraphics[width=1\textwidth]{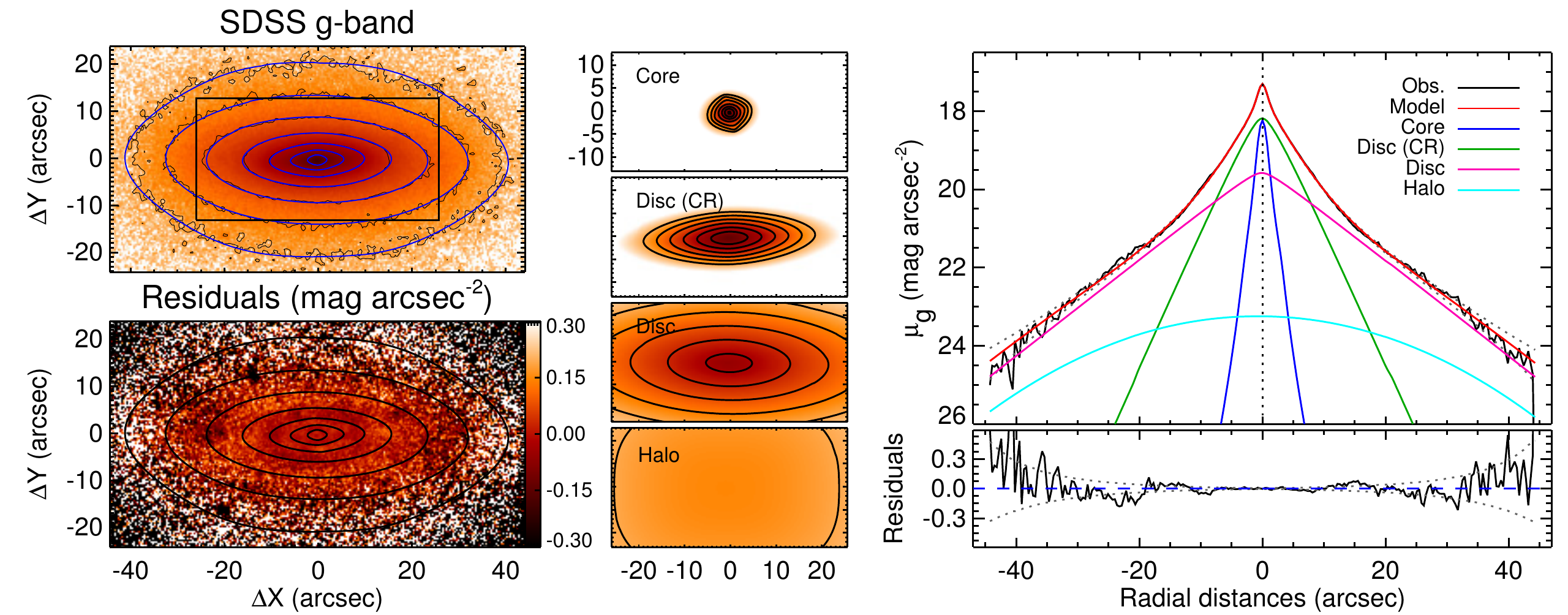} 
\caption{The 2D photometric decomposition results of the SDSS $g$-band image. The
top panel in the left block presents the galaxy image superimposed by contours
according to the surface brightness of 18, 19, 20, 21, 22, 23, 24 mag~arcsec$^ {-2}$. 
Blue lines correspond to the contours of the model image. The bottom
panel in the left block presents a residual map in magnitudes with the model
contours as a reference. The middle block shows images of the model subcomponents
(from top to bottom): core, disc (CR), main disc, halo. The right block presents a
one-dimensional light profile along the major axis of the galaxy (black line)
with the overplotted total 4-component model (in red) and the subcomponent
profiles (in colours). Dotted lines show the level of uncertainties. The bottom
panel in this block corresponds to the fitting residuals.}\label{fig_ph_decomp}
\end{figure*}

\begin{figure}
\centering
\includegraphics[width=0.45\textwidth]{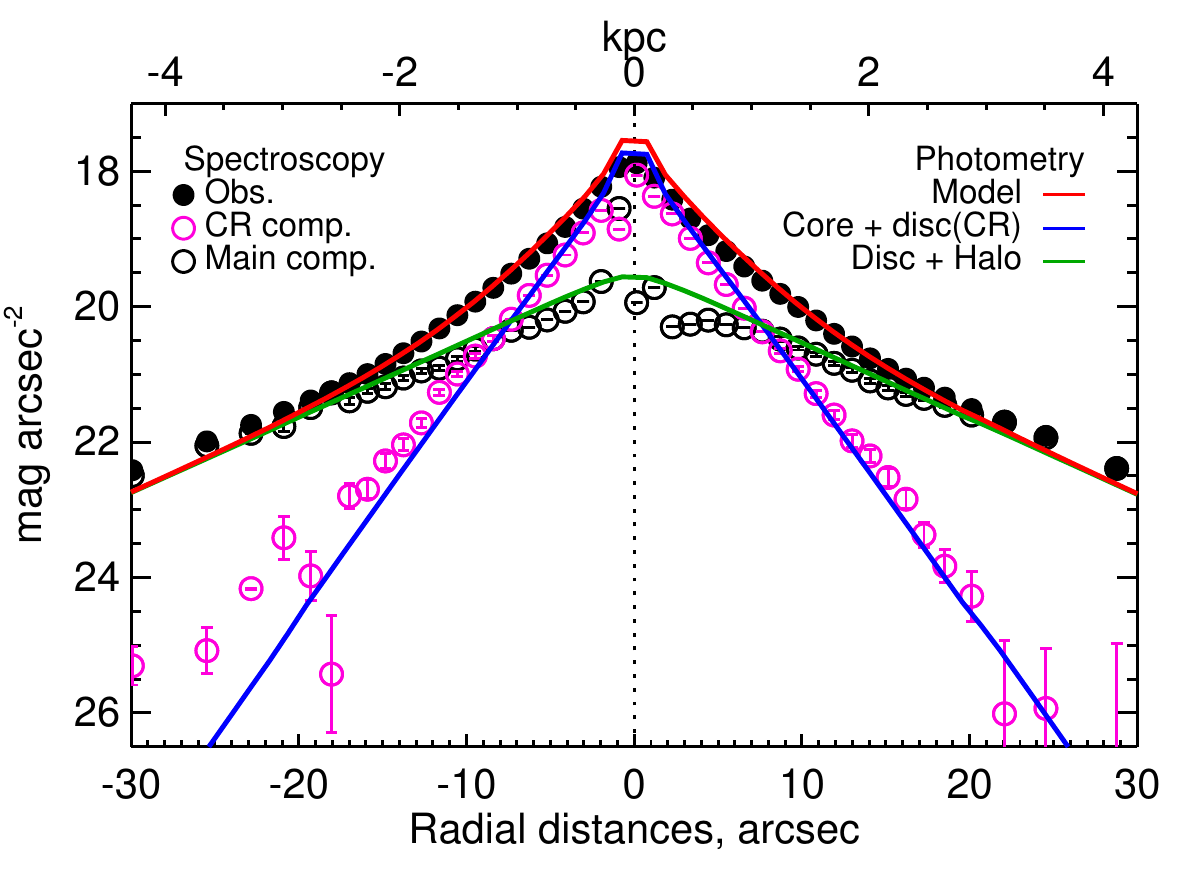} 
\caption{Comparison of the photometric and spectroscopic decompositions.
Filled circles show the light profile extracted from our long-slit spectrum,
open circles are the results of the spectroscopic decomposition on the main (in black) and
counter-rotating (in magenta) components. Solid lines display the result of the
photometric decomposition. The red solid line corresponds to the total photometric
model, the blue solid line is a sum of disc (CR) and core components
that is assumed to be associated with the counter-rotating component, the green solid
line is a main component represented by the sum of the main disc and halo
components. Here we took into account the atmospheric seeing difference
between SCORPIO observations (FWHM=1.5~arcsec) and SDSS $g$-band image
(FWHM=1.0~arcsec) as well as sampling effects. This comparison clearly demonstrates
the successful agreement between spectroscopic and photometric approach.}
\label{fig_decomp_comp}
\end{figure}

Finally we adopted a four-component model that consists of a S\'ersic nuclear
core, two exponential discs, and a S\'ersic halo. Hereafter, the more centrally
concentrated exponential disc is referred to as ``CR'' assuming its association
with the counter-rotating stellar component. Fig.~\ref{fig_ph_decomp}
graphically demonstrates the results of the decomposition, and
Fig.~\ref{fig_decomp_comp} displays the comparison between the spectroscopic and
photometric results. Table~\ref{tbl_galfit_results} presents the best-fitting
parameters of the photometric model as well as parameter uncertainties.
We estimated the parameter uncertainties from the covariance matrix as a
standard output of the Levenberg--Marquardt minimization algorithm used in the
\textsc{galfit} routine and also by Monte-Carlo simulations for a hundred
realizations of synthetic galaxy images in the same manner as for the spectroscopic
decomposition. Error estimates from both approaches are in good agreement.
We emphasize that our uncertainty estimates should probably be considered as lower limits of the true
parameter uncertainties \citep{IMFIT,GEMS} and should be used with caution.

\begin{table*}
\caption{Structural component parameters of NGC~448 from the photometric
decomposition. Column (2) corresponds to the central surface brightness $\mu_0$,
(3) is the disc scalelength $h$ or effective radius $r_e$ for S\'ersic
component, (4) is the S\'ersic index $n$, (5) is the positional angle $PA$, (6)
is the axis ratio $q=b/a$, (5) is the diskiness/boxiness $C_0$, (6) is the
contribution to the total image ($C/T$). The parameter uncertainties
outcome from \textsc{galfit} routine, while errors in the parentheses are
extracted by means of Monte-Carlo simulations.}

% Parameter uncertainties correspond to
% the formally computed statistical errors by \textsc{galfit} during
% $\chi^2$-minimization.
\label{tbl_galfit_results}
\centering
\resizebox{\textwidth}{!}{%
\begin{tabular}{llllllll}
\hline
Name & $\mu_0$ & $h$, $r_e$ & $n$ & $q$ & $PA$ & $C_0$ & $C/T$\\
          & (mag/arcsec$^2$) & (arcsec) & & & (deg) & &\\
(1) & (2) & (3) & (4) & (5) & (6) & (7) & (8)\\
\hline
% \hline
% Core & $ 17.25 \pm 0.21(0.25) $ & $  0.86 \pm 0.05(0.11) $ & $  1.86 \pm 0.14(0.19) $ & $  0.718 \pm 0.010(0.013) $ & $ 122.40 \pm  1.40(1.13) $ & --  &    0.05 \\
% Disc$_{CR}$ & $ 19.75 \pm 0.01(0.02) $ & $  3.14 \pm 0.01(0.02) $ & --  & $  0.280 \pm 0.002(0.04) $ & $ 118.31 \pm  0.03(0.05) $ & $   0.20 \pm  0.02(0.03) $ &    0.27 \\
% Disc & $ 21.68 \pm 0.01(0.01) $ & $  8.94 \pm 0.05(0.05) $ & --  & $  0.367 \pm 0.001(0.002) $ & $ 115.08 \pm  0.04(0.04) $ & $   0.04 \pm  0.01(0.01) $ &    0.50 \\
% Halo & $ 25.49 \pm 0.06(0.07) $ & $ 24.63 \pm 0.27(0.20) $ & $  0.55 \pm 0.01(0.02) $ & $  0.785 \pm 0.008(0.010) $ & $ 115.86 \pm  0.70(0.59) $ & $   0.79 \pm  0.07(0.10) $ &    0.18 \\
% \hline
\hline
% After updated sky subtraction
Core & $ 17.25 \pm 0.21(0.25) $ & $  0.86 \pm 0.05(0.11) $ & $  1.86 \pm 0.14(0.19) $ & $  0.724 \pm 0.010(0.013) $ & $ 122.37 \pm  1.43(1.13) $ & --  &    0.05 \\
Disc$_{CR}$ & $ 19.75 \pm 0.01(0.02) $ & $  3.13 \pm 0.01(0.02) $ & --  & $  0.279 \pm 0.002(0.04) $ & $ 118.33 \pm  0.03(0.05) $ & $   0.19 \pm  0.02(0.03) $ &    0.27 \\
Disc & $ 21.68 \pm 0.01(0.01) $ & $  8.84 \pm 0.05(0.05) $ & --  & $  0.368 \pm 0.001(0.002) $ & $ 115.09 \pm  0.04(0.04) $ & $   0.05 \pm  0.01(0.01) $ &    0.51 \\
Halo & $ 25.56 \pm 0.06(0.07) $ & $ 24.21 \pm 0.27(0.20) $ & $  0.49 \pm 0.01(0.02) $ & $  0.766 \pm 0.008(0.010) $ & $ 116.02 \pm  0.70(0.59) $ & $   0.71 \pm  0.07(0.10) $ &    0.17 \\

\hline	
\end{tabular}}
\end{table*}

\subsection{Results}

\begin{figure}
\centering
\includegraphics[width=0.45\textwidth]{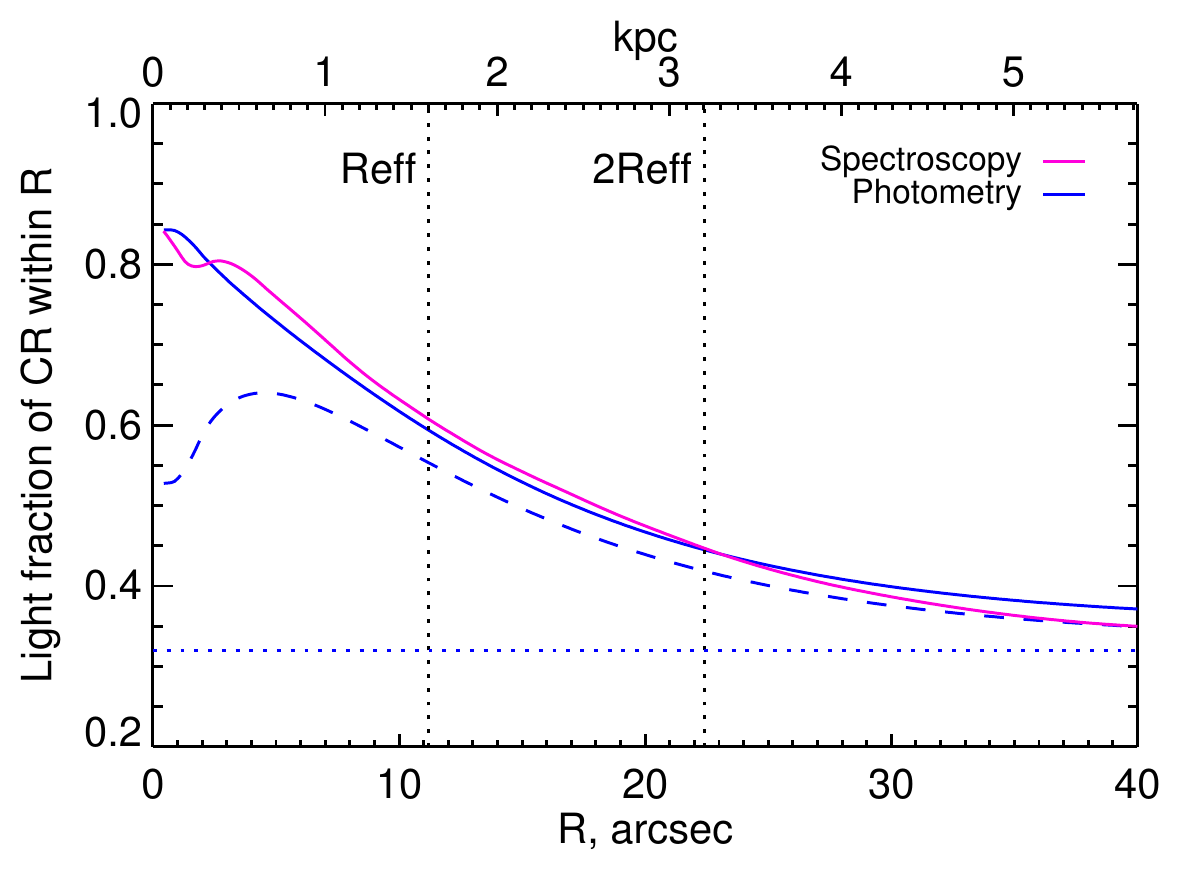} 
\caption{
Total light fraction of the \textit{counter-rotating} component (CR) within
the galactocentric (not projected) radius $R$, which is calculated based on
the light profiles extracted along the \textit{major axis} from the long-slit
spectroscopic decomposition (magenta) and from the two-dimensional photometric
decomposition (blue). The dashed blue line corresponds to the fraction of the Disc$_{CR}$
component only, while the solid blue line shows Disc$_{CR}$ + Core. The blue dotted
horizontal line presents a total contribution of the CR component (Disc$_{CR}$ + Core)
to the galaxy image  from the \textsc{galfit} modeling. Note that we
calculated the light fraction from  one-dimensional light profiles assuming an
axisymmetric galaxy light distribution.\label{fig_light_fraction}}
\end{figure}

It is clearly seen on the position-velocity diagram 
(Fig.~\ref{fig_nonpar_losvd}) as well as on the radial profiles of the light
contribution of the individual components extracted from the long-slit data (the right
panel of Fig.~\ref{fig_sp_decomp_profiles}) that the counter-rotating component
dominates in the integrated spectrum in the central region of NGC~448. The total light
fraction of the counter-rotating disc within one effective radius 
($R_{eff}=11.2$~arcsec) is 60~per~cent and decreases to 45~per~cent within $2R_{eff}$
(see Fig.~\ref{fig_light_fraction}).

Our photometric results completely agree with the spectral decomposition.
Fig.~\ref{fig_decomp_comp} demonstrates the comparison of the
major axis light profiles extracted from the spectrum and from the image. Note, that in
the central region we cannot separate the contributions of the kinematical
components due to the very small difference of their line-of-sight velocities.
Therefore, it is more reasonable to compare the total contribution of the CR disc
\textit{and} the core component together. Fig.~\ref{fig_light_fraction} shows the
comparison of the total light fraction of the \textit{counter-rotating} component within
a given galactocentric radius.

% because,
% according to it, the disc (CR) component contributes  27~per~cent to the total
% galaxy light (see Table~\ref{tbl_galfit_results}). It is more correct to compare
% a total contribution of the CR disc \textit{and} the core component that
% together account for 32~per~cent of the light, is because in the central region
% we cannot separate the contributions of the kinematical components due to the
% very small difference of the line-of-sight velocities.

We assume that the counter-rotating component has a disc
morphology. This assumption is supported by a relatively low velocity dispersion
($V/\sigma \sim 2$) and by the exponential shape of the light profile distribution
(see the right panel of Fig.~\ref {fig_sp_decomp_profiles}). Our 
two-dimensional photometric model of NGC~448 contains two exponential discs. In
our photometric model, the counter-rotating disc (CR) has a significant diskiness
of isophotes which is characterized by $C_0=0.2$. The nearly edge-on disc orientation
can produce such values of the $C_0$ coefficient. The \textsc{galfit}
package contains the edge-on disc model. We tested it and found that it
yields higher values of $\chi^2$ as well as the statistical criteria AIC and
BIC which may be connected to the disc thickness changing along the radius.

We cannot determine the exact orientation of the counter-rotating disc with
respect to the main disc by relying only on our long-slit kinematical data
because we observed NGC~448 only in one slit position. However, there are
several arguments which support the discs to be settled in the same
plane: (i) the equal amplitudes of the line-of-sight velocity variations along
the radius; (ii) the co-alignment of the kinematical  major axes of the central
counter-rotating region and the outermost main disc component  (see \atlas\
velocity maps in \citet{atlas3d_ii}); (iii) similar values of the isophote major
axis position angles and inclinations\footnote{Indeed, assuming flatnesses of
the discs the axis ratios $q_{CR}=0.280$ and $q=0.367$ produce inclinations
$i_{CR}=\arccos(q_{CR})=74^\circ$ and $i=68^\circ$, respectively}.

We found that in the galaxy center both discs contain old stars ($T_{SSP}\approx
9$ Gyr) having sub-solar metallicity putting our results in agreement with the
literature \citep{caldwell2003,atlas3d_xxx}. Our deep spectroscopic data allowed
us to determine a noticeable age gradient $\Delta\log T_{CR}=-0.087\pm0.026$
dex~per~dex\footnote{The gradient calculated as a slope of a linear fit of the
form $a + b \log(r/R_{eff})$. The uncertainties of age and metallicity
measurements are invoked to the $\chi^2$ minimization routine to estimate
1-$\sigma$ error of the gradient.} in the counter-rotating stellar disc, while in the
main stellar disc the age profile is flat ($\Delta\log T_{MD}=-0.009\pm0.024$
dex~per~dex). The stellar metallicity gradients are also noticeably
different: $\Delta\log Z_{CR}=-0.09\pm0.04$ dex~per~dex while $\Delta\log
Z_{MD}=-0.33\pm0.05$ dex per dex.

We have subtracted the best-fitting absorption-line model from the observed spectrum
in order to obtain a pure emission-line spectrum. Our long-slit spectrum covers
the positions of the H$\beta$ and [O{\sc iii}] emission lines. We have not
detected any significant signal in these lines.

\begin{figure}
\includegraphics[width=\linewidth]{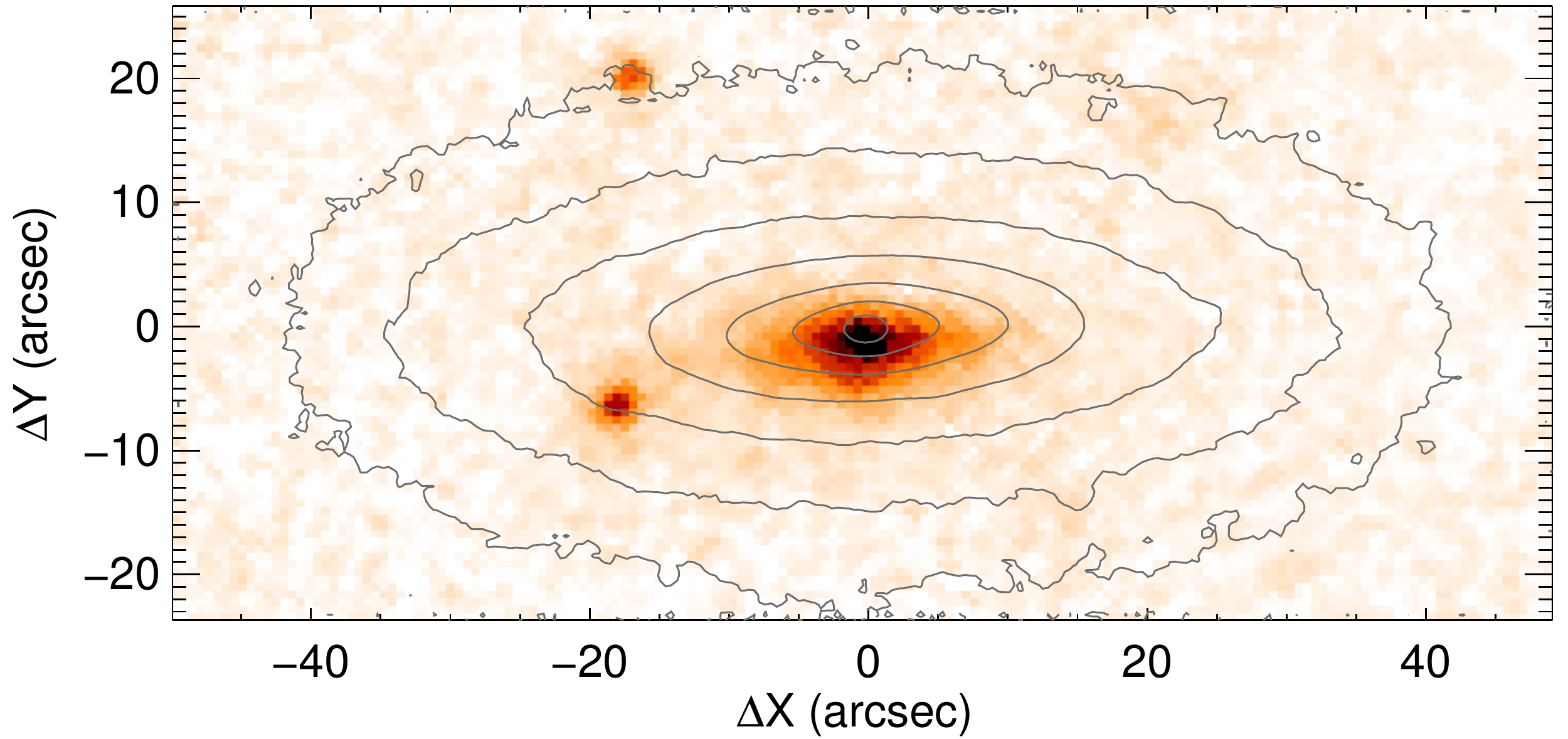}
\caption{
A stellar continuum subtracted \textit{Spitzer} image at 8$\mu$ with overlaid
contours according to the surface brightness values of 18, 19, 20, 21, 22, 23, 24
mag~arcsec$^{-2}$. Some non-stellar signal is seen in the central part, $R<10^{\prime\prime}$. }\label{fig:spitzer}
\end{figure}

Although we have not clearly detected emission lines in the spectra of NGC~448, we
found signs of possible star formation in infrared (IR) images. We analysed
available archival \textit{Spitzer} images of the galaxy at 3.6$\mu$ and 8$\mu$.
While emission in the first band originates almost exclusively from the stellar
light, the dominate sources of emission at  8$\mu$ are
polycyclic aromatic hydrocarbons (PAH), being excited by UV photons from massive and/or hot
stars. Hence, if bright emission is seen in an 8$\mu$ image after the subtraction of the
stellar population contribution, this should indicate the presence of gas excited
either by the star formation in a galaxy or by hot evolved blue horizontal branch stars. 
In order to subtract the stellar continuum in the 8$\mu$ image, we used the
calibration by \cite{Ciesla14} that estimated the 8$\mu$ stellar contribution
to be about 24.9 per cent of the total flux at 3.6$\mu$ for early-type galaxies. 
The subtracted image corresponding to the non-stellar emission at 8$\mu$ is shown in
Fig.~\ref{fig:spitzer} with overlaid contours from the optical SDSS-$g$ image.

We calculated the stellar mass of the counter-rotating component by using mass-to-light ratios
($M/L$) determined from spectral fitting based on the \textsc{pegase.hr} models
with the Salpeter IMF and the light profile which we recovered from the
photometry. As we described above the counter-rotating component is
assumed to be associated with both the core and the CR disc components in the
photometric decomposition. Since the core is very centrally concentrated, we
used a single stellar mass-to-light ratio of $(M/L_g)_{core}=6.3$ that
corresponds to the central stellar population properties. For the disc (CR) component, we
estimated $(M/L_g)_ {disc}$ pixel-by-pixel by using the gradients of the stellar
population parameters  retrieved from the long-slit spectrum. The 
total mass of the counter-rotating component is excpected to be
$M_{CR}=9.0^{+2.7}_{-1.8}\cdot10^{9}M_\odot$
including $M_{disc}=7.1^{+2.6}_{-1.8}\cdot10^ 9M_\odot$ and
$M_{core}=1.8^{+0.2}_{-0.2}\cdot10^{9}M_\odot$. Mass uncertainties have been
calculated by using uncertainties of the stellar population properties.

\section{Discussion and Conclusions} 

The main challenge in the studies of galaxies possessing counter-rotating
components is to establish a possible source of material with a different
direction of the angular momentum and to clarify at least some details of
the counter-rotating disc formation.

It is generally accepted that the presence of counter-rotating stars within the
main stellar disc is a result of external material acquisition
\citep{Corsini_review}. Nevertheless, \citet{EvansCollett94} suggested a
scenario of internal origin of counter-rotating stars where stars take
retrograde orbits during  the bar dissolution process (separatrix crossing).
From the stellar population point of view, only counter-rotating stellar discs
with identical stellar population properties can be produced  in the framework
of this scenario. Hence, we decline it in the case of NGC~448 because the two
discs have significantly different stellar population properties.
Moreover, \citet{EvansCollett94} suggested the separatrix crossing as a natural 
mechanism for building identical (with the same ages and scale lengths) 
counter-rotating discs in NGC~4550. However, the applicability of such mechanism for 
counter-rotating discs with very different scale lengths is not obvious.

Many recent detailed studies of disc galaxies with large-scale 
counter-rotating components support the external origin of 
counter-rotating stars. In all studied galaxies, stellar population properties
derived from the spectra play an important
role. It has been shown that the counter-rotating components detected in NGC~3593,
NGC~5719, NGC~4191, NGC~4550, and IC~719
\citep{coccato_n3593_n4550,coccato_n5719,Johnston_n4550, coccato_n4191,ic719}
have younger stellar populations compared to the main stellar discs, and
their ionized gas rotates in the same direction as the secondary stellar
components, i.e. it also counter-rotates with  respect to the main disc. These
findings favour the scenario where counter-rotating  stars have been formed
\textit{in-situ} from the externally accreted gas.

Cosmological filaments and gas-rich satellites are considered as main candidate
sources of external cold gas. However, it is often difficult to unambiguously
disentangle between them. In both scenarios, one can expect
to find either metal-poor or metal-rich counter-rotating stellar population
with respect to the main disc, depending  on the star formation history while
the brief duration of the subsequent star formation event results in the $\alpha$-element
enhancement. The diversity of properties is observed. For instance,  NGC~3593
and NGC~5719 indicate lower stellar metallicity in their counter-rotating discs
with respect to the main discs while for NGC~4550 and NGC~4191 both discs have
similar populations; and the counter-rotating stars in IC~719 and NGC~448 are
more metal-rich than their main stellar discs. The duration of the accretion
events can be also various that results in the super-solar $\alpha$-element to
iron ratios in the counter-rotating stellar populations for short timescales
while prolonged formation  of a counter-rotating component should produce a
solar $\alpha$/Fe abundance ratio and/or significant age gradient.

Nevertheless, for some targets the scenario of accretion from gas-rich
neighbouring galaxies looks more plausible. For instance, no doubt that NGC~5719
and IC~719 have accreted their counter-rotating stellar and gaseous discs from
the closest neighbours because of the H{\sc i} bridge \citep{Vergani2007}
between the galaxy and its neighbour, NGC~5713, in the case of NGC~5719, and the common H{\sc i}
envelope \citep{Grossi2009} including both galaxies, in the case of IC~719 and its
neighbour IC~718.

Recently, \citet{coccato_n4191} have investigated the structure of the Virgo cluster
S0 galaxy NGC~4191 by using the IFU spectroscopy obtained with the
VIRUS-W spectrograph.  They have interpreted the results of photometric as well
as spectroscopic decomposition of NGC~4191 in the context of the recent
cosmological simulation by \citet{Algorry2014} where it is demonstrated
how distinct cosmological filamentary structures providing material with the
opposite spins can finally produce two coplanar counter-rotating stellar discs.
\citet{coccato_n4191} found significant negative age gradients in both
components of NGC~4191  correlating at least in the central 20~arcsec that
indicates the inside-out formation of both components.

In our case of NGC~448, we observe a different situation. First, we show that
the two independent approaches, the spectral and the photometric 2D
decompositions, are in perfect agreement. The secondary counter-rotating disc
reveals a detectable negative age gradient ($\Delta\log T_{CR}=-0.087\pm0.026$
dex per dex) that also gives an evidence for a prolonged inside-out formation
process during approximately $3\dots4$ Gyrs. At the same time, the main stellar
disc does not show any significant age gradient ($\Delta\log
T_{MD}=-0.009\pm0.024$ dex per dex)  that indicates that it was probably formed
independently, by another mechanism than the secondary one. This is also
supported by different metallicity gradients ($\Delta\log
Z_{CR}=-0.09\pm0.04$ dex~per~dex, $\Delta\log Z_{MD}=-0.33\pm0.05$ dex per
dex). Moreover, based on our stellar metallicity measurements, we suggest that
the accreted gas was pre-enriched by metals in the companion galaxy and that the
scenario of the counter-rotating disc formation from cosmological filaments is
implausible for NGC~448.

Taking into account the large mass of the counter-rotating stellar population 
$M_{CR}=9.0^{+2.7}_{-1.8}\cdot10^{9}M_\odot$ and indistinguishable properties
of the stellar populations in the central region of the galaxy, we conclude that a
significant fraction of accreted material had been already transformed into
stars prior to the moment when the disc was acquired by NGC~448. All these 
observational facts indicate that the most probable formation scenario for
the  counter-rotating disc in NGC~448 is a merger event
with a consequent prolonged gaseous accretion.

Remarkably, neither the search for emission lines of ionized gas in our optical
spectra, nor attempts of radio observation of CO emission \citep{atlas3d_iv} and
H{\sc i} 21-cm line yielded any detection of gas in NGC~448. Only IR
image indirectly points to the probable presence of gas. The formation of its
counter-rotating disc occurred approximately $6\dots7$ Gyrs ago since the
remnants of the gas could be swept away by a recent tidal interaction between
the galaxy and its satellite  GALEXMSC J011516.31-013456.8 supported by the
presence of a low surface brightness bridge connecting the two galaxies.

In conclusion, several recent studies, including ours, demonstrate that
various scenarios to form counter-rotating stellar components can take place in
real galaxies. It is important to expand the sample of disc galaxies with
counter-rotation studied in detail, in order to produce quantitative
conclusions on the probability of different formation scenarios supported by the
statistics. The on-going large spectroscopic surveys of galaxies such as CALIFA
\citep{califa}, MANGA \citep{manga}, SAMI \citep{sami} would help to compile a
list of good candidates for subsequent detailed studies. However, 
dedicated follow-up deep spectroscopic observations (IFU or long-slit) 
are required in order to obtain high signal-to-noise ratios and also
the appropriate spectral resolution in order for the spectral decomposition
techniques to be successful.

\section*{Acknowledgments}

I.K. thanks Lodovico Coccato and Sergey Khoperskov for useful discussions.
The work was supported by the Russian Science Foundation project 14-22-00041
``VOLGA -- A View On the Life of GAlaxies''. The final interpretation of the
data and paper writing was performed during visits to Chamonix workshop and The
Research Institute in Astrophysics and Planetology in Toulouse which were
supported by MD-7355.2015.2, RFBR 15-32-21062, and RFBR-CNRS 15-52-15050 grants.
The project used computational resources funded by the M.V.~Lomonosov Moscow
State University Program of Development. The Russian 6-m telescope is exploited
under the financial support by the Russian Federation Ministry of Education and
Science (agreement No14.619.21.0004, project ID RFMEFI61914X0004).  This
research has made use of the NASA/IPAC Extragalactic Database (NED) which is
operated by the Jet Propulsion Laboratory, California Institute of Technology,
under contract with the National Aeronautics and Space Administration, and of
the Lyon Extragalactic Database (LEDA). In this study, we used the SDSS DR12
data. Funding for the SDSS and SDSS-II has been provided by the Alfred P. Sloan
Foundation, the Participating Institutions, the National Science Foundation, the
U.S. Department of Energy, the National Aeronautics and Space Administration,
the Japanese Monbukagakusho, the Max Planck Society, and the Higher Education
Funding Council for England. The SDSS Web site is http://www.sdss.org/.

%%%%%%%%%%%%%%%%%%%%%%%%%%%%%%%%%%%%%%%%%%%%%%%%%%

%%%%%%%%%%%%%%%%%%%% REFERENCES %%%%%%%%%%%%%%%%%%

% The best way to enter references is to use BibTeX:

\bibliographystyle{mnras}
\bibliography{refs} % if your bibtex file is called example.bib

% Alternatively you could enter them by hand, like this:
% This method is tedious and prone to error if you have lots of references
% \begin{thebibliography}{99}
% \bibitem[\protect\citeauthoryear{Author}{2012}]{Author2012}
% Author A.~N., 2013, Journal of Improbable Astronomy, 1, 1
% \bibitem[\protect\citeauthoryear{Others}{2013}]{Others2013}
% Others S., 2012, Journal of Interesting Stuff, 17, 198
% \end{thebibliography}

%%%%%%%%%%%%%%%%%%%%%%%%%%%%%%%%%%%%%%%%%%%%%%%%%%

% %%%%%%%%%%%%%%%%% APPENDICES %%%%%%%%%%%%%%%%%%%%%

% \appendix

% \section{Some extra material}

% If you want to present additional material which would interrupt the flow of the main paper,
% it can be placed in an Appendix which appears after the list of references.

% %%%%%%%%%%%%%%%%%%%%%%%%%%%%%%%%%%%%%%%%%%%%%%%%%%

% Don't change these lines
\bsp	% typesetting comment
\label{lastpage}
\end{document}